\documentclass[aps,preprint,groupedaddress,a4paper]{revtex4-2}

\usepackage[T1]{fontenc}   
\usepackage{textcomp} 
\usepackage{graphicx}

\usepackage{mathtools}
\usepackage[colorlinks = true,linkcolor = blue,urlcolor  = blue,citecolor = blue,anchorcolor = blue]{hyperref}
\usepackage{xcolor, soul}
\usepackage{color}

\usepackage{makecell}

\DeclarePairedDelimiter\ket{\lvert}{\rangle}
\DeclarePairedDelimiter\bra{\langle}{\rvert}

\begin{document}


\title{Qutrit-Native Spatial-Orbital Encoding for Resource-Efficient Quantum Chemistry Simulation}


\author{Sumin Lim}
\email[]{suminlim@kaist.ac.kr}
\affiliation{Graduate School of Quantum Science and Technology, KAIST, Daejeon, 34141, Republic of Korea}



\date{\today}

\begin{abstract}
Additional levels in a qudit can serve as more than extra computational capacity and can represent physically relevant correlations when the algorithmic design reflects the underlying physics of the problem. We demonstrate this principle for quantum chemistry simulation using a qutrit-native spatial-orbital encoding, where $\ket{0}$, $\ket{1}$, and $\ket{2}$ denote orbital occupation states. We establish a qutrit simulation framework that combines number-conserving pair-transfer and broken-pair generators with a projected-Hamiltonian energy estimator constructed from qutrit populations and coherences. Benchmarks for H$_2$, LiH, and H$_2$O show that the encoding reproduces the relevant potential-energy curves, while retaining a compact quantum resource structure. Direct comparison with the pair-only restriction confirms that the third level reduces errors by tens of mHa in LiH and by more than $100~\mathrm{mHa}$ in H$_2$O. Despite a partial spin-coupling truncation in the compact H$_2$O encoding, the resulting energy curve deviates from FCI by only a few mHa over the tested range. Compared with a conventional qubit-based UCCSD approach, the qutrit scheme requires half as many quantum units and reduces the scaling of the parameterized ansatz generator count from $\mathcal{O}(M^4)$ to $\mathcal{O}(M^2)$, where $M$ is the number of spatial orbitals. This work elucidates how a physically motivated qutrit encoding can reduce quantum-resource requirements while retaining controlled electronic-structure accuracy, thereby providing a concrete starting point for broader qudit-native quantum algorithm design.         

\end{abstract}


\maketitle

\section{Introduction}

Accurate electronic-structure calculations are central to molecular and materials science, but their computational cost increases rapidly with system size. In a finite spin-orbital basis, the full configuration-interaction (FCI) space grows combinatorially with the numbers of electrons and orbitals, making exact classical diagonalization impractical beyond modest active spaces \cite{knowles1984new, sherrill1999configuration}.
Quantum algorithms can provide a natural framework for representing many-body wavefunctions \cite{aspuru2005simulated,peruzzo2014variational,kandala2017hardware} and offer a complementary route to correlated electronic structure. Quantum phase estimation \cite{nielsen2010quantum} can in principle estimate eigenenergies of a molecular Hamiltonian to arbitrary accuracy, but its practical implementation requires deep quantum circuits and fault-tolerant quantum hardware. In the absence of such hardware, the variational quantum eigensolver (VQE) has been extensively investigated as an alternative approach to molecular electronic-structure  calculations on near-term quantum devices. However, even in the VQE approach, scaling up quantum circuits while maintaining sufficiently low errors remains challenging and motivates the investigation of resource-efficient quantum algorithms designed for current quantum hardware \cite{preskill2018quantum}. Under these circumstances, reducing quantum register size and circuit resources remains an important part of algorithm design. 

Most quantum chemistry simulation methods start from a spin-orbital representation and map fermionic occupations to qubits through transformations such as Jordan--Wigner \cite{jordan1928paulische} or Bravyi--Kitaev \cite{bravyi2002fermionic}. This construction retains the full spin-orbital occupation information, but requires two qubits for each spatial orbital. For closed-shell and near-closed-shell systems, more compact approaches that assign one qubit to each spatial orbital \cite{kim2024variational} have been proposed. Such seniority-zero methods restrict the accessible space to paired configurations, as in doubly occupied configuration interaction (DOCI), and therefore exclude broken-pair and higher-seniority contributions \cite{henderson2014seniority, wahlen2018influence}, which can be recovered through extended ansatz terms or additional heuristic corrections \cite{kim2024variational}. This tradeoff motivates representations that preserve the compact spatial-orbital picture while explicitly including singly occupied configurations.

Feynman's seminal proposal for quantum simulation emphasized the direct simulation of quantum systems using quantum-mechanical degrees of freedom \cite{feynman2018simulating}. From this point of view, native multilevel quantum systems, or qudits, can offer an alternative route to resource-efficient quantum algorithm design \cite{macdonell2021analog, chizzini2024qudit, kiktenko2025colloquium}. In recent years, several works have explored qudits for quantum chemistry simulation \cite{cao2024emulating, kim2024qudit, wang2025efficient} and other quantum-computing applications \cite{shi2026efficient, fischer2023universal}. Some of these approaches encode the computational space of $N$ qubits into a single $2^N$-dimensional qudit. Here, we instead investigate an encoding that associates qutrit levels directly with spatial-orbital occupation states for quantum chemistry simulation. Each active spatial orbital is represented by one qutrit, where $\ket{0}$, $\ket{1}$, and $\ket{2}$ denote empty, singly occupied, and doubly occupied states, respectively. Restricting the local space to $\ket{0}$ and $\ket{2}$ recovers the occupation structure of a pair-only description, whereas the added $\ket{1}$ level introduces singly occupied, broken-pair configurations directly into the variational Hilbert space. The three qutrit levels therefore carry a direct chemical interpretation rather than serving only to enlarge the computational space.

To realize this encoding, we construct number-conserving qutrit operations for pair transfer and broken-pair generation, project conventional active-space Hamiltonians into the qutrit occupation basis, and formulate an energy estimator based on occupation and coherence measurements. We benchmark the framework using H$_2$, LiH, and H$_2$O as systems of increasing complexity. H$_2$ validates the two-qutrit pair-transfer scheme, LiH tests a compact three-qutrit active-space model with singly occupied configurations, and H$_2$O tests the six-qutrit occupation encoding in a larger multi-electron active space. We also compare the full construction with its pair-only restriction, examine performance under finite-error conditions, and compare its register and ansatz resources with those of a conventional qubit-based UCCSD reference. These analyses characterize the electronic-structure accuracy, correlation recovery, finite-error performance, and quantum-resource requirements of the qutrit-native construction.

Beyond quantum chemistry, qudit-based approaches have been explored for their potential advantages in algorithm design, resource efficiency, and quantum information processing across a broad range of applications. Examples include qudit-based quantum error correction and fault-tolerant schemes \cite{lim2023fault, mezzadri2024fault, lim2025demonstrating, debry2026error}, simulations of elementary-particle dynamics such as three-flavor neutrino oscillations \cite{turro2025qutrit, spagnoli2025collective}, emergent phenomena in condensed-matter systems \cite{ticea2025observation}, high-dimensional quantum communication \cite{cozzolino2019high, cerf2002security, bechmann2000quantum}, and quantum information processing \cite{sawaya2020resource, jankovic2024noisy}. These examples illustrate that higher-dimensional degrees of freedom can be incorporated directly into algorithm design rather than used merely as extra capacity for encoding binary information. Recent work toward fault-tolerant logical qudits \cite{uy2025qudit, lim2026fault} further motivates the native use of multilevel systems. Within this context, extending the present qutrit construction to more general $d$-dimensional qudits may provide further opportunities for resource-efficient quantum simulation.

\section{Qutrit-native encoding, circuit structures, and energy estimation}
\label{sec:qutrit_method}

\begin{figure}[!htbp]
\includegraphics[width=16cm]{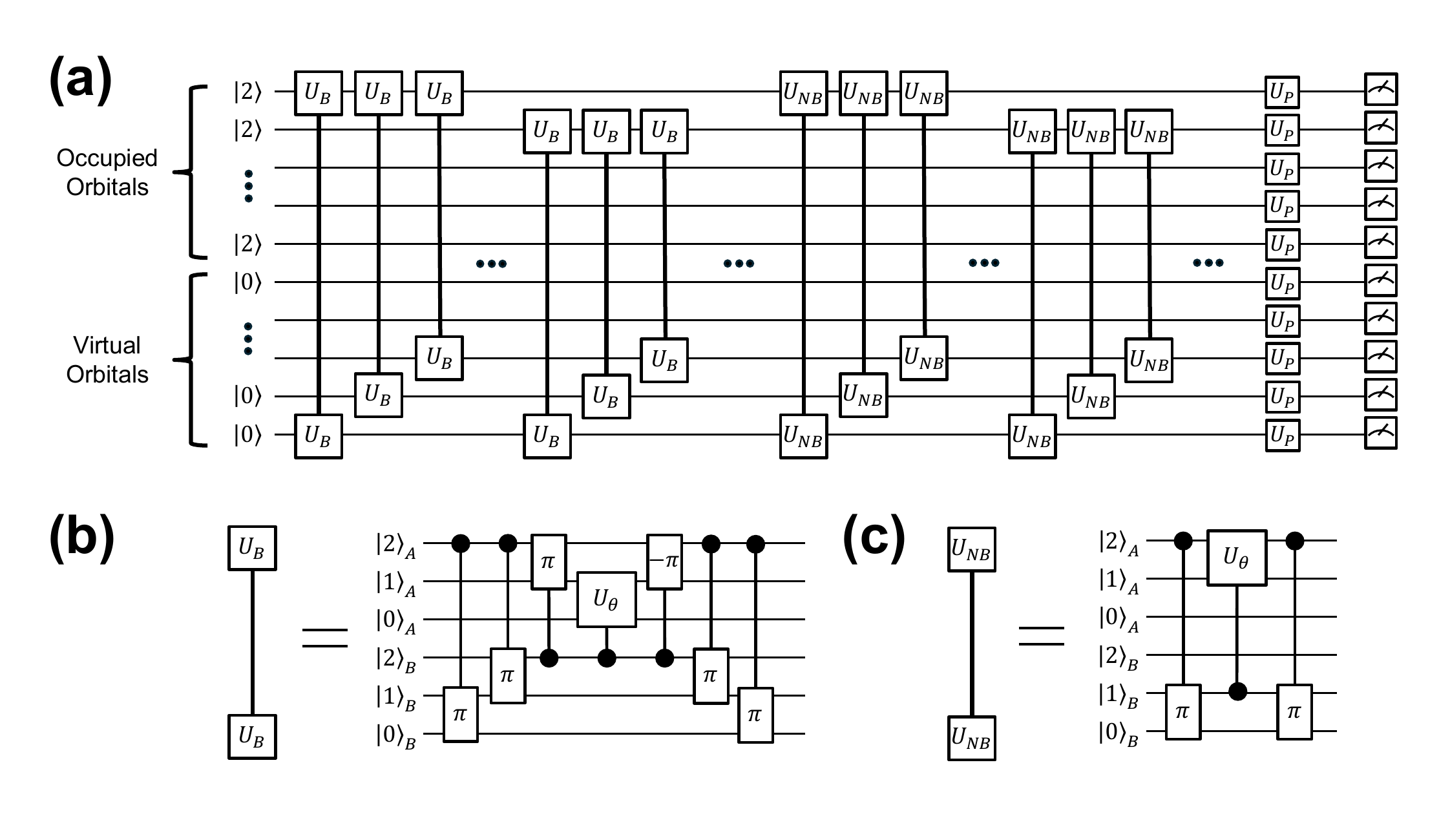}
\caption{ Overview of the qutrit-native variational quantum eigensolver. (a) A Hartree--Fock occupation state is prepared using one qutrit per active spatial orbital, followed by pair-transfer blocks $U_B$, broken-pair blocks $U_{\mathrm{NB}}$, and measurement-basis rotations $U_P$. (b) Pair-transfer block $U_B$ implementing $\ket{20}_{ij}\leftrightarrow\ket{02}_{ij}$, together with an example decomposition. (c) Broken-pair block $U_{\mathrm{NB}}$ implementing $\ket{20}_{ij}\leftrightarrow\ket{11}_{ij}$, together with an example decomposition.}
\label{fig:qutrit_workflow}
\end{figure}

Figure~\ref{fig:qutrit_workflow}(a) summarizes the overall workflow of the qutrit-native VQE simulation
introduced in this work. A molecular Hamiltonian is first constructed in a selected active space using a classical
electronic-structure calculation. After mapping each active spatial orbital to one qutrit and preparing the Hartree--Fock reference state, a number-conserving variational ansatz state is generated using the pair-transfer and broken-pair rotations shown in Figure~\ref{fig:qutrit_workflow} (b) and (c). The energy is then evaluated from occupation probabilities and coherences between encoded configurations, as detailed in Sections~\ref{subsec:qutrit_hamiltonian} and  ~\ref{subsec:qutrit_measurement_examples}. 

\subsection{Spatial-orbital occupation encoding}
\label{subsec:qutrit_encoding}

Let $M$ denote the number of active spatial orbitals and $N_e$ the number of
active electrons. Instead of assigning one qubit to each spin orbital as in conventional spin-orbital encodings, we
assign one qutrit to each spatial orbital $p$, with the local states
\begin{equation}
\begin{aligned}
\ket{0}_p &\equiv \text{empty spatial orbital }p,\\
\ket{1}_p &\equiv \text{singly occupied spatial orbital }p,\\
\ket{2}_p &\equiv \text{doubly occupied spatial orbital }p.
\end{aligned}
\label{eq:local_qutrit_encoding}
\end{equation}
The fixed-electron-number occupation space is therefore
\begin{equation}
\mathcal Q_{N_e,M}
=
\left\{
\ket{\mathbf n}_q
=
\ket{n_1n_2\cdots n_M}_q:
n_p\in\{0,1,2\},\;
\sum_{p=1}^{M}n_p=N_e
\right\}.
\label{eq:qutrit_occupation_space}
\end{equation}
For the closed-shell systems considered here, the Hartree--Fock reference is represented by assigning $\ket{2}$ to the
doubly occupied active orbitals and $\ket{0}$ to the virtual orbitals. For example, after freezing the oxygen $1s$-like core orbital, the Hartree--Fock reference for the H$_2$O active space used below is $\ket{222200}_q$.

For configurations containing two singly occupied orbitals, we denote the two orbitals by $i$ and $j$, with $i<j$, and associate the qutrit occupation $\ket{\cdots 1_i \cdots 1_j \cdots}_q$ with the normalized spin-adapted open-shell singlet
\begin{equation}
\ket{S_{ij}}
=
\frac{1}{\sqrt{2}}
\left(
\ket{\alpha_i\beta_j}
-
\ket{\beta_i\alpha_j}
\right).
\label{eq:two_open_shell_singlet}
\end{equation}
This definition fixes the relative phase convention used for the projected-Hamiltonian matrix elements below.

For occupation strings containing four singly occupied orbitals, as encountered in the H$_2$O active space considered below, two linearly independent singlet configuration-state functions can correspond to the same spatial-orbital occupation string. In the compact encoding used here, we apply a single singlet-coupling criterion to each such occupation string. We label the four singly occupied orbitals as $p<q<r<s$ according to increasing molecular-orbital energy and select the $(pq)(rs)$ singlet coupling. The consequences of this truncation and coupling choice are quantified in Appendix~\ref{SI_chapter_B}.

\subsection{Number-conserving qutrit ansatz}
\label{subsec:qutrit_ansatz}

The variational state used in the simulation is prepared from the encoded Hartree--Fock reference as
\begin{equation}
\ket{\psi(\boldsymbol\theta)}_q
=
U(\boldsymbol\theta)
\ket{\mathbf n_{\mathrm{HF}}}_q .
\label{eq:qutrit_variational_state}
\end{equation}

The ansatz contains two types of effective two-qutrit rotations. The pair-transfer block $U_B^{ij}(\theta)$, corresponding to a bosonic excitation, acts nontrivially in the two-qutrit subspace spanned by $\{\ket{20}_{ij},\ket{02}_{ij}\}$ and transfers an
electron pair between spatial orbitals $i$ and $j$,
\begin{equation}
\ket{20}_{ij}
\leftrightarrow
\ket{02}_{ij}.
\label{eq:UB_transition}
\end{equation}
Restricting every local qutrit to $\{\ket{0},\ket{2}\}$ and using only $U_B$ corresponds to the pair-only limit analyzed in
Sec.~\ref{qutrit_native_level_effect}.

The second type is the broken-pair block $U_{\mathrm{NB}}^{ij}(\theta)$, which describes a non-bosonic excitation and acts nontrivially in the two-qutrit subspace spanned by $\{\ket{20}_{ij},\ket{11}_{ij}\}$,
\begin{equation}
\ket{20}_{ij}
\leftrightarrow
\ket{11}_{ij},
\label{eq:UNB_transition}
\end{equation}
thereby activating the singly occupied qutrit level and introducing a singlet open-shell configuration. Reversing the orbital ordering provides the complementary $\ket{02}_{ij}\leftrightarrow\ket{11}_{ij}$ channel. Both blocks conserve the total electron number within their respective logical transition subspaces.

The full ansatz is an ordered product of these blocks,
\begin{equation}
U(\boldsymbol\theta)
=
\prod_{\ell}
U_{\mathrm{NB}}^{i_\ell j_\ell}(\theta_\ell)
\prod_{k}
U_B^{i_kj_k}(\theta_k),
\label{eq:qutrit_ansatz_product}
\end{equation}

where $\boldsymbol\theta$ is the full variational parameter vector containing all pair-transfer and broken-pair rotation parameters. The blocks acting on overlapping orbital pairs need not commute. In the present numerical benchmarks, we use a simple and systematic ordering in which the $U_B$ sweep is applied first, followed by the $U_{\mathrm{NB}}$ sweep, with the descending orbital-pair order illustrated in Fig.~\ref{fig:qutrit_workflow}(a). This ordering is a practical choice rather than a fixed requirement of the encoding. Depending on the available computational resources, the $U_B$ and $U_{\mathrm{NB}}$ sweeps can be repeated to increase variational flexibility, and their sequence can also be rearranged according to the molecular structure or hardware constraints.

With the encoded configuration space and the variational circuit defined, we next construct the molecular Hamiltonian projected onto this space and formulate how its expectation value can be obtained from qutrit measurements.

\subsection{Projected molecular Hamiltonian and energy evaluation}
\label{subsec:qutrit_hamiltonian}

In a finite spatial-orbital basis, the active-space molecular Hamiltonian used in this work is written as
\begin{equation}
\begin{aligned}
H_{\mathrm{AS}}
={}&
E_0
+
\sum_{pq\in A}\sum_{\sigma}
h_{pq}
a_{p\sigma}^{\dagger}a_{q\sigma}\\
&+
\frac{1}{2}
\sum_{pqrs\in A}\sum_{\sigma\tau}
(pq|rs)
a_{p\sigma}^{\dagger}
a_{r\tau}^{\dagger}
a_{s\tau}
a_{q\sigma},
\end{aligned}
\label{eq:active_space_hamiltonian}
\end{equation}
where $A$ denotes the selected active spatial orbitals, $\sigma,\tau\in\{\alpha,\beta\}$ label spin, $h_{pq}$ denotes the effective one-electron integrals, and $(pq|rs)$ denotes the two-electron integrals in chemists' notation. $E_0$ collects all scalar energy contributions, including the nuclear-repulsion energy and any fixed frozen-core contribution. For frozen-core calculations, the core--active Coulomb and exchange contributions are included in the effective one-electron integrals $h_{pq}$.

For each qutrit occupation string $\ket{\mu}_q$, let $\ket{\Phi_\mu}$ denote the corresponding normalized determinant or the spin-adapted configuration-state function defined above. The molecular Hamiltonian in the qutrit occupation basis is then
\begin{equation}
H_Q
=
\sum_{\mu,\nu\in\mathcal Q_{N_e,M}}
(H_Q)_{\mu\nu}
\ket{\mu}_q {}_q\bra{\nu},
\qquad
(H_Q)_{\mu\nu}
=
\bra{\Phi_\mu}
H_{\mathrm{AS}}
\ket{\Phi_\nu}.
\label{eq:qutrit_projected_hamiltonian}
\end{equation}
The same construction is used for all molecular examples. The one- and two-electron integrals are calculated classically, and the matrix elements are evaluated using the Slater--Condon rules in the selected spin-adapted basis. 

The qutrit variational state can then be expanded as
\begin{equation}
\ket{\psi(\boldsymbol\theta)}_q
=
\sum_{\mu}
c_{\mu}(\boldsymbol\theta)\ket{\mu}_q,
\label{eq:qutrit_state_expansion}
\end{equation}

and, for a real molecular Hamiltonian, the energy is evaluated as 

\begin{equation}
\begin{aligned}
E(\boldsymbol\theta)
=&
{}_q\bra{\psi(\boldsymbol\theta)}
H_Q
\ket{\psi(\boldsymbol\theta)}_q\\
=&
\sum_{\mu}
(H_Q)_{\mu\mu}
P_{\mu}(\boldsymbol\theta)
+
\sum_{\mu<\nu}
(H_Q)_{\mu\nu}
C_{\mu\nu}(\boldsymbol\theta),
\end{aligned}
\label{eq:qutrit_energy_estimator}
\end{equation}
where
\begin{equation}
P_\mu(\boldsymbol\theta)
=
|c_\mu(\boldsymbol\theta)|^2,
\qquad
C_{\mu\nu}(\boldsymbol\theta)
=
2\,\mathrm{Re}
\left(c_\mu^*(\boldsymbol\theta)c_\nu(\boldsymbol\theta)\right).
\label{eq:qutrit_population_coherence}
\end{equation}
The diagonal populations $P_\mu$ are obtained directly from qutrit occupation-basis readout. The off-diagonal coherences are obtained from joint occupation correlations after applying appropriate local basis rotations $U_P$ to each orbital whose occupation differs between $\mu$ and $\nu$. 

For two distinct qutrit levels $a,b\in\{0,1,2\}$, we define local Pauli-like operators acting within the selected two-level subspace,
\begin{equation}
X_p^{ab}
=
\ket{a}_p {}_p\!\bra{b}
+
\ket{b}_p {}_p\!\bra{a},
\qquad
Y_p^{ab}
=
-i\ket{a}_p {}_p\!\bra{b}
+
i\ket{b}_p {}_p\!\bra{a}.
\label{eq:qutrit_transition_operators}
\end{equation}

The local basis rotations $U_P$, chosen for the transitions being measured, map the corresponding $X$- or $Y$-type coherence onto joint qutrit occupation correlations, which are then measured in the occupation basis.

\subsection{Representative measurement channels and electronic coefficients}
\label{subsec:qutrit_measurement_examples}

The following examples illustrate how measured qutrit coherences and their electronic coefficients enter the molecular energy estimator. Compact integral expressions are given for a two-active-electron singlet prototype, which directly describes the LiH encoded basis. For general many-electron active spaces, the energy is evaluated from the complete matrix elements 
$(H_Q)_{\mu\nu}$ together with the populations and coherences in Eqs.~\eqref{eq:qutrit_energy_estimator} and~\eqref{eq:qutrit_population_coherence}.

\paragraph{Double--double channel.}

A double--double (D--D) transition transfers an electron pair,
\begin{equation}
\ket{D_i}_q
=
\ket{\cdots2_i\cdots0_j\cdots}_q
\leftrightarrow
\ket{D_j}_q
=
\ket{\cdots0_i\cdots2_j\cdots}_q .
\end{equation}

The measurement involves the $0\leftrightarrow2$ transitions on qutrits $i$ and $j$. Using the Pauli-like operators defined above, the corresponding real coherence operator is

\begin{equation}
\mathcal C_{ij}^{\mathrm{D-D}}
=
\frac{1}{2}
\left(
X_i^{02}X_j^{02}
+
Y_i^{02}Y_j^{02}
\right),
\label{eq:DD_measurement_operator}
\end{equation}

whose expectation value gives the coherence $C_{D_iD_j}$. The corresponding Slater--Condon matrix element is

\begin{equation}
(H_Q)_{D_iD_j}
=
\bra{\Phi_{D_i}}
H_{\mathrm{AS}}
\ket{\Phi_{D_j}}
=
(ij|ji)
\equiv
K_{ij},
\label{eq:DD_coefficient}
\end{equation}

where \(K_{ij}\) is the exchange, or pair-hopping, integral. Thus this contribution to Eq.~\eqref{eq:qutrit_energy_estimator} is

\begin{equation}
E_{ij}^{\mathrm{D-D}}
=
K_{ij}
\left\langle
\mathcal C_{ij}^{\mathrm{D-D}}
\right\rangle .
\end{equation}

\paragraph{Double--single channel with a shared orbital.}

A double--single (D--S) transition connects a doubly occupied configuration to an open-shell singlet,
\begin{equation}
\ket{D_i}_q
=
\ket{\cdots2_i\cdots0_j\cdots}_q
\leftrightarrow
\ket{S_{ij}}_q
=
\ket{\cdots1_i\cdots1_j\cdots}_q .
\end{equation}

The measurement combines the $1\leftrightarrow2$ transition on qutrit $i$ and the $0\leftrightarrow1$ transition on qutrit $j$. The corresponding real coherence operator is

\begin{equation}
\mathcal C_{ij}^{\mathrm{D-S}}
=
\frac{1}{2}
\left(
X_i^{12}X_j^{01}
+
Y_i^{12}Y_j^{01}
\right),
\label{eq:DS_shared_measurement_operator}
\end{equation}

whose expectation value gives the coherence $C_{D_iS_{ij}}$.

For the normalized two-electron singlet of Eq.~\eqref{eq:two_open_shell_singlet}, the corresponding Slater--Condon matrix element is

\begin{equation}
\begin{aligned}
(H_Q)_{D_i,S_{ij}}
&=
\bra{\Phi_{D_i}}
H_{\mathrm{AS}}
\ket{\Phi_{S_{ij}}}\\
&=
\frac{1}{\sqrt{2}}
\left[
2h_{ij}
+
(ii|ij)
+
(ij|ii)
\right].
\end{aligned}
\label{eq:DS_shared_coefficient}
\end{equation}

The factor $1/\sqrt{2}$ follows from the normalization of the spin-adapted open-shell state. Thus this contribution to Eq.~\eqref{eq:qutrit_energy_estimator} is

\begin{equation}
E_{ij}^{\mathrm{D-S}}
=
\frac{1}{\sqrt{2}}
\left[
2h_{ij}
+
(ii|ij)
+
(ij|ii)
\right]
\left\langle
\mathcal C_{ij}^{\mathrm{D-S}}
\right\rangle .
\end{equation}

\paragraph{Double--single channel without a shared orbital.}
When the initial doubly occupied orbital \(i\) is distinct from both orbitals of the final singlet pair,

\begin{equation}
\ket{D_i}_q
=
\ket{\cdots2_i\cdots0_j\cdots0_k\cdots}_q
\leftrightarrow
\ket{S_{jk}}_q
=
\ket{\cdots0_i\cdots1_j\cdots1_k\cdots}_q ,
\qquad
i\notin\{j,k\}.
\end{equation}

Three qutrit occupations change:
\(2\leftrightarrow0\) on \(i\) and \(0\leftrightarrow1\) on \(j\) and \(k\). The corresponding coherence is obtained from three-site joint correlators constructed from the $0\leftrightarrow2$ transition on qutrit $i$ and the
$0\leftrightarrow1$ transitions on qutrits $j$ and $k$, using the corresponding $X$- and $Y$-type measurement settings. In the two-electron prototype,
\begin{equation}
(H_Q)_{D_i,S_{jk}}
=
\frac{1}{\sqrt{2}}
\left[
(ij|ik)
+
(ik|ij)
\right].
\label{eq:DS_nonshared_coefficient}
\end{equation}

This matrix element multiplies the corresponding measured coherence $C_{D_iS_{jk}}$ in Eq.~\eqref{eq:qutrit_energy_estimator}.

\paragraph{Single--single channel with a shared orbital.}

The qutrit configurations corresponding to two open-shell singlets sharing orbital $i$ are connected as
\begin{equation}
\ket{S_{ij}}_q
\leftrightarrow
\ket{S_{ik}}_q .
\end{equation}
At the qutrit occupation level, the single occupation moves from $j$ to $k$, so the measurement involves the $0\leftrightarrow1$ transitions on qutrits $j$ and $k$. The corresponding real coherence operator is
\begin{equation}
\mathcal C_{jk}^{\mathrm{S-S}}
=
\frac{1}{2}
\left(
X_j^{01}X_k^{01}
+
Y_j^{01}Y_k^{01}
\right),
\label{eq:SS_shared_measurement_operator}
\end{equation}
whose expectation value gives the coherence $C_{S_{ij}S_{ik}}$ after resolving the shared occupation $n_i=1$ in the occupation readout.

For the two-electron prototype, the corresponding Slater--Condon matrix element is
\begin{equation}
\begin{aligned}
(H_Q)_{S_{ij},S_{ik}}
&=
\bra{\Phi_{S_{ij}}}
H_{\mathrm{AS}}
\ket{\Phi_{S_{ik}}}\\
&=
h_{jk}
+
\frac{1}{2}
\left[
(ii|jk)
+
(ik|ji)
+
(ji|ik)
+
(jk|ii)
\right].
\end{aligned}
\label{eq:SS_shared_coefficient}
\end{equation}
This matrix element multiplies the measured coherence
$C_{S_{ij}S_{ik}}$ in Eq.~\eqref{eq:qutrit_energy_estimator}.

More general D--S or S--S coherences involving additional changed orbitals are treated in the same way, with the corresponding local measurement settings applied to each changed qutrit and the electronic coefficients
evaluated directly from Eq.~\eqref{eq:qutrit_projected_hamiltonian}.

\section{Results}
\label{sec:results}
\subsection{H$_2$ Model}

\begin{figure}[!htbp]
\includegraphics[width=16cm]{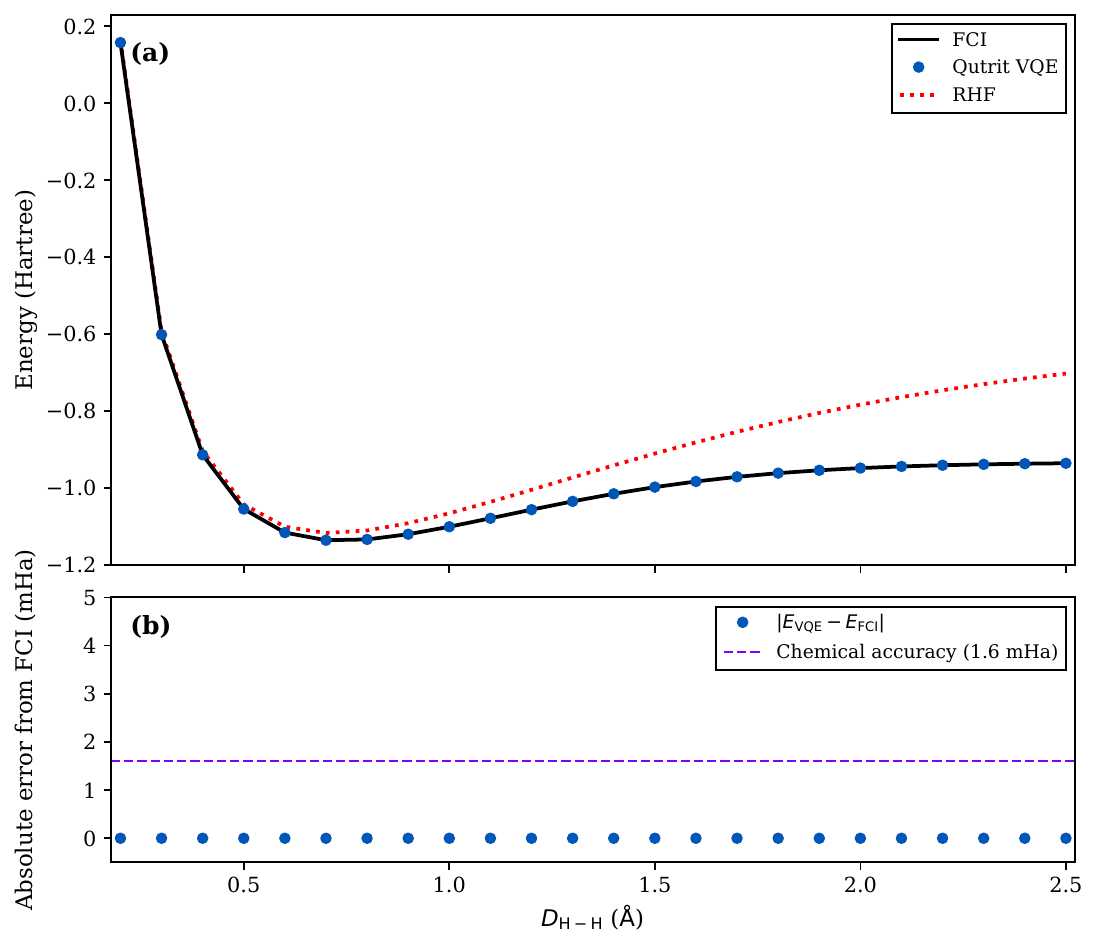}
\caption{Ground-state potential-energy curve of H$_2$ in the STO-3G basis using two qutrits. (a) RHF, FCI, and qutrit-VQE energies as functions of the H--H distance. (b) Absolute qutrit-VQE energy error relative to FCI. The
horizontal dashed line marks the chemical-accuracy scale, $1~\mathrm{kcal\,mol^{-1}}=1.5936~\mathrm{mHa}$.}
\label{H2_sim_qutrit}
\end{figure}

We first use H$_2$ as a basic validation of the qutrit mapping, pair-transfer ansatz, Hamiltonian projection, and energy-evaluation procedure introduced above. In the STO-3G basis, the two-electron active space of H$_2$ consists of the bonding and antibonding spatial orbitals, which are represented by two qutrits. Within the ground-state symmetry sector considered here, the only relevant correlation arises from electron-pair transfer between these two orbitals \cite{szabo2012modern}. The Hartree--Fock reference for qutrits $A$ and $B$, $\ket{2_A0_B}_q$, is therefore connected to $\ket{0_A2_B}_q$ by a single $U_B$ rotation, and the projected Hamiltonian reduces to this paired two-configuration subspace.

Figure~\ref{H2_sim_qutrit}(a) compares the qutrit-VQE potential-energy curve with restricted Hartree--Fock (RHF) and FCI over distances from $0.2$ to $2.5~\text{\AA}$. The optimized one-parameter qutrit circuit agrees with FCI to numerical precision at every geometry. The absolute VQE--FCI deviations shown in Fig.~\ref{H2_sim_qutrit}(b) are far below the conventional chemical-accuracy scale ($1~\mathrm{kcal\,mol^{-1}}=1.5936~\mathrm{mHa}$)~\cite{feller2007probing} and reflect numerical precision rather than a physically meaningful error. Within this minimal H$_2$ model and under ideal quantum circuit conditions, the agreement therefore verifies the consistency of the qutrit encoding, the pair-transfer block $U_B$, and the projected-energy construction. The effects of finite-fidelity state preparation, qutrit operations, and readout, together with finite-shot sampling, are examined separately for the same H$_2$ benchmark in Sec.~\ref{Sec:Implementation_and_error_benchmark}.

\subsection{LiH Model}

\begin{figure}[!htbp]
\includegraphics[width=16cm]{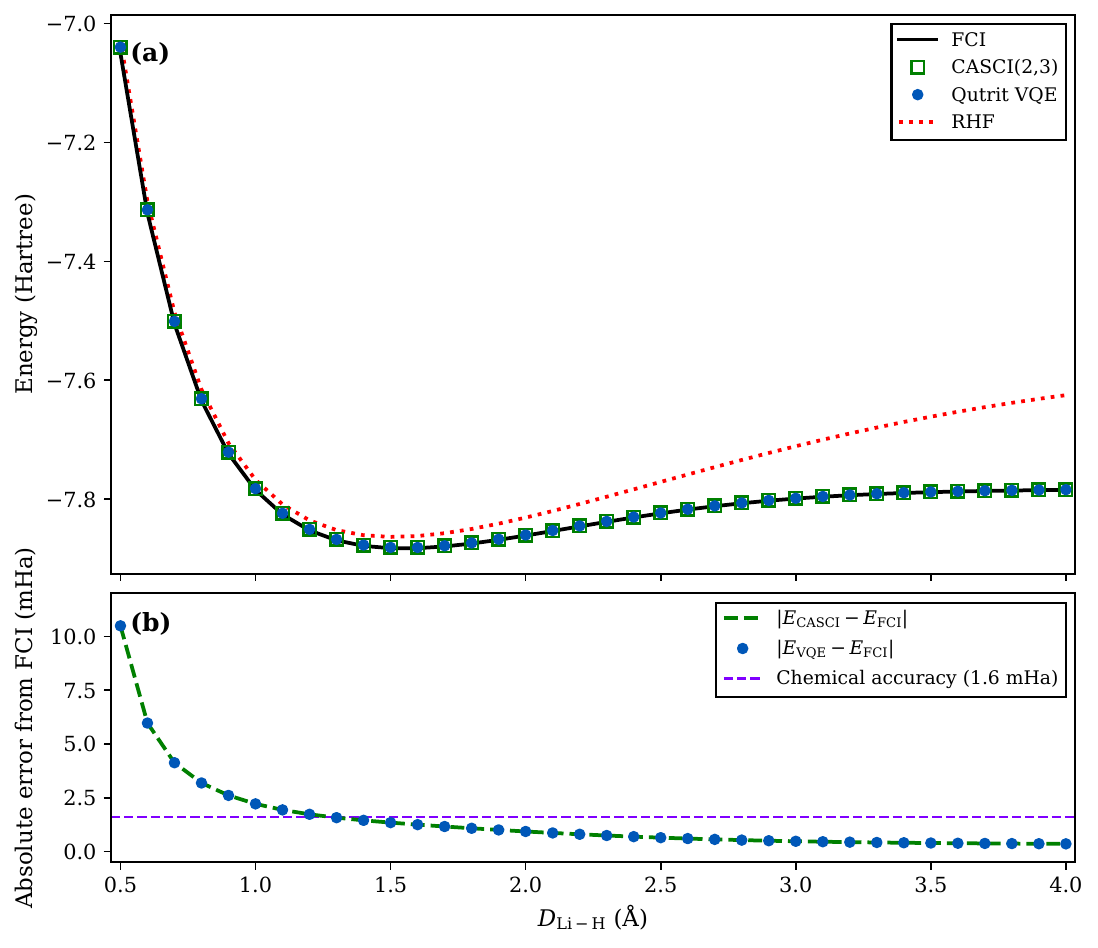}
\caption{Ground-state potential-energy curve of LiH in the STO-3G basis using the selected frozen-core CASCI(2,3) active space and three qutrits. (a) RHF, FCI, CASCI(2,3), and qutrit-VQE energies as functions of the Li--H distance.
(b) Absolute CASCI(2,3)--FCI and qutrit-VQE--FCI energy deviations. The horizontal dashed line marks the chemical-accuracy scale.}
\label{LiH_sim_qutrit}
\end{figure}

LiH provides the first benchmark in which the singly occupied qutrit level, $\ket{1}$, participates explicitly in the variational space. In the STO-3G minimal basis, LiH is described by six spatial molecular orbitals. After freezing the lowest Li $1s$-like molecular orbital, the frozen-core model retains the two valence electrons in the five remaining spatial orbitals, yielding a complete active space configuration interaction (CASCI) calculation, denoted CASCI(2,5)~\cite{roos1980complete}. Although this model provides a reference energy very close to FCI, its direct occupation encoding would require five qutrits. Because a central goal of the qutrit framework is to balance electronic-structure accuracy against quantum-resource cost, we instead use three spatial orbitals with dominant H $1s$, Li $2s$, and Li $2p_x$ character, where the $x$ direction is chosen along the Li--H bond. The resulting CASCI(2,3) model requires only three qutrits. This reduction introduces a modest active-space truncation error, but the selected three-orbital model remains accurate in the chemically relevant region and provides the most favorable fixed three-orbital choice over most of the sampled geometries. The orbital-selection procedure and its comparison with CASCI(2,5) and the other CASCI(2,3) choices are detailed in Appendix~\ref{SI_chapter_A}.

Within the fixed two-electron subspace, the three-qutrit register contains the six encoded singlet configurations
\begin{equation}
\mathcal Q_{2,3}
=\{\ket{200}_q, \ket{020}_q, \ket{002}_q, \ket{110}_q, \ket{101}_q, \ket{011}_q\}.
\end{equation}
The first three are paired configurations, whereas the last three contain two singly occupied orbitals with the two electrons coupled to a singlet. Starting from the Hartree--Fock reference $\ket{200}_q$, the ansatz uses three pair-transfer blocks $U_B$ and three broken-pair blocks $U_{\mathrm{NB}}$, requiring six variational parameters. The six basis configurations are represented within a three-qutrit register rather than the six-qubit register required by a conventional spin-orbital occupation encoding. They span the complete singlet CASCI(2,3) space. 

Figure~\ref{LiH_sim_qutrit}(a) compares the qutrit-VQE potential-energy curve with the RHF, FCI, and selected CASCI(2,3) results. On the energy scale of Figure~\ref{LiH_sim_qutrit}(a), both the selected CASCI(2,3) and qutrit-VQE curves closely follow the FCI reference throughout the calculated bond-length range. The absolute errors relative to FCI are shown in Fig.~\ref{LiH_sim_qutrit}(b). At the sampled potential-energy minimum, $D_{\mathrm{Li-H}}=1.5~\text{\AA}$, the CASCI(2,3) error relative to FCI is approximately $1.35~\mathrm{mHa}$, whereas the additional VQE residual relative to CASCI(2,3) is negligible on this scale. The maximum VQE--CASCI(2,3) gap remains on the order of $10^{-7}~\mathrm{mHa}$, showing that the combined $U_B+U_{\mathrm{NB}}$ ansatz can access the correlated configurations required to reproduce the CASCI(2,3) reference. The larger total deviations occur mainly in the strongly compressed region, where the three-orbital active-space approximation becomes less accurate. Thus, near the potential-energy minimum, the qutrit VQE essentially reaches the CASCI(2,3) limit, leaving the three-orbital active-space truncation as the dominant source of error relative to FCI.

\subsection{H$_2$O Model}

\begin{figure}[!htbp]
\includegraphics[width=16cm]{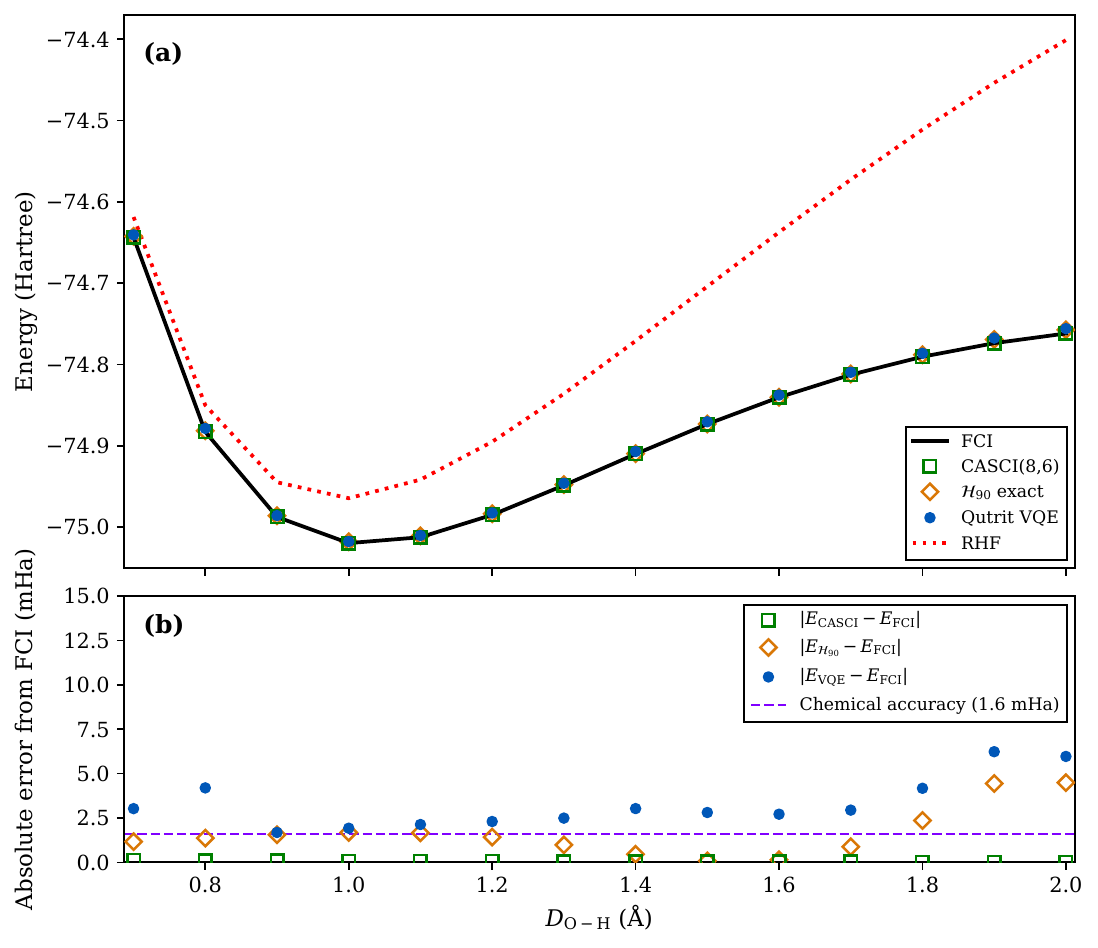}
\caption{Ground-state potential-energy curve of H$_2$O for symmetric O--H stretching in the STO-3G basis. (a) RHF, FCI, frozen-core CASCI(8,6), exact diagonalization within the selected 90-state space $\mathcal{H}_{90}$, and six-qutrit VQE energies as functions of the common O--H distance. (b) Absolute energy deviations from FCI for CASCI(8,6), exact $\mathcal{H}_{90}$, and qutrit VQE. The horizontal dashed line marks the chemical-accuracy scale.}
\label{H2O_sim_qutrit}
\end{figure}

We next examine symmetric O--H stretching of H$_2$O over the bond-length range  $D_{\mathrm{O-H}}=0.7$--$2.0~\text{\AA}$ at a fixed H--O--H angle of $104.5^\circ$. This system provides a more demanding test of the qutrit framework than H$_2$ or LiH because several paired and broken-pair configurations contribute simultaneously in a larger multi-electron active space. In the STO-3G basis, the O $1s$-like molecular orbital is frozen, leaving eight active electrons in six spatial orbitals, each mapped to one qutrit. Starting from the Hartree--Fock occupation state $\ket{222200}_q$, the ansatz contains 15 pair-transfer blocks $U_B$ and 30 broken-pair blocks $U_{\mathrm{NB}}$, giving 45 variational parameters.

The corresponding fixed-electron-number occupation space is
\begin{equation}
\mathcal{Q}_{8,6}
=
\left\{
\ket{n_1n_2\cdots n_6}_q:
n_i\in\{0,1,2\},\;
\sum_i n_i=8
\right\},
\end{equation}
and contains 90 qutrit occupation strings.  This 90-dimensional space differs from the complete CASCI(8,6) singlet space, which contains 105 spin-adapted configuration-state functions. For each occupation pattern with four singly occupied orbitals, the present encoding retains one singlet-coupling function and omits the second independent coupling. We denote the corresponding 90-dimensional spin-adapted subspace by $\mathcal H_{90}$. The $(pq)(rs)$ singlet-pairing rule defined in Sec.~\ref{subsec:qutrit_encoding} and the associated 15-dimensional reduction are analyzed quantitatively in Appendix~\ref{SI_chapter_B}.

Figure~\ref{H2O_sim_qutrit}(a) shows the potential-energy curves obtained from RHF, FCI, CASCI(8,6), exact diagonalization within $\mathcal H_{90}$, and six-qutrit VQE. At this scale, the CASCI(8,6), exact $\mathcal H_{90}$, and qutrit VQE results remain close to FCI across the bond-length range, while the RHF curve deviates increasingly toward larger O--H distances. The corresponding absolute errors relative to FCI are presented in Figure~\ref{H2O_sim_qutrit}(b). The CASCI(8,6) curve differs from FCI by only $0.0109$--$0.1181~\mathrm{mHa}$ across the sampled geometries, showing that the frozen-core active-space error is small relative to both the chemical-accuracy scale and the other error contributions. Exact diagonalization within $\mathcal H_{90}$ gives a total error relative to FCI of $0.0850$--$4.4861~\mathrm{mHa}$. The additional $\mathcal H_{90}$--CASCI(8,6) difference ranges from $0.0496$ to $4.4752~\mathrm{mHa}$ and increases primarily as the O--H bonds are stretched. By construction, exact diagonalization within $\mathcal H_{90}$ gives the lowest energy available within the selected qutrit-encoded space.

The optimized six-qutrit VQE reproduces the overall shape of the FCI potential-energy curve, including the minimum at $D_{\mathrm{O-H}}=1.0~\text{\AA}$. Using a multi-start COBYLA optimization within a finite desktop-scale budget, the calculation gives a total VQE--FCI error of $1.68$--$6.23~\mathrm{mHa}$ across the bond-length range. The remaining VQE gap relative to the exact $\mathcal H_{90}$ reference is $0.125$--$2.83~\mathrm{mHa}$. Together with the exact-space results discussed above, these values establish the overall scales of the error sources. The active-space truncation contributes less than $0.12~\mathrm{mHa}$, while both the spin-coupling truncation and the remaining VQE residual reach the few-mHa level.

The present H$_2$O calculation applies each of $U_B$ and $U_\mathrm{NB}$ sweeps only once. Repeating these sweeps can enlarge the set of states accessible by the ansatz and may further lower the optimized energy, at the cost of more parameterized quantum gate blocks and a larger classical optimization budget. Overall, the H$_2$O benchmark achieves several-mHa agreement with FCI using a compact six-qutrit circuit with 45 parameterized gate blocks.

\section{Qutrit-native correlations beyond pair-only encoding}
\label{qutrit_native_level_effect}

\begin{figure}[!htbp]
\includegraphics[width=16cm]{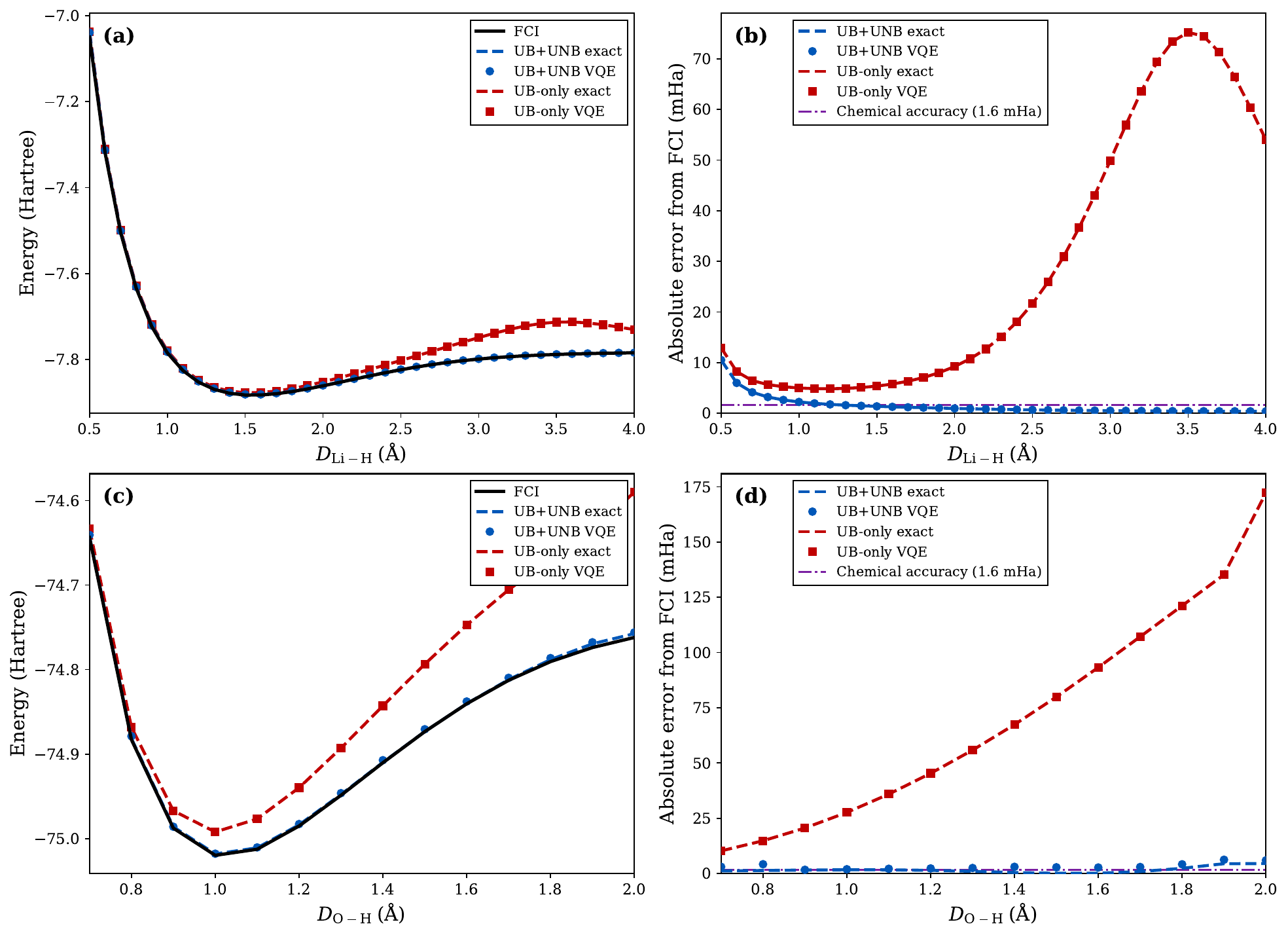}
\caption{Effect of the qutrit-native singly occupied level on the LiH and H$_2$O potential-energy calculations. (a),(b) LiH potential-energy curves and absolute FCI errors for exact and VQE calculations in the $U_B+U_{\mathrm{NB}}$ and $U_B$-only spaces. (c),(d) Corresponding results for H$_2$O. The $U_B$-only space restricts each spatial orbital to $\ket{0}$ and $\ket{2}$, whereas $U_B+U_{\mathrm{NB}}$ additionally includes the singly occupied $\ket{1}$ state. The horizontal dash-dotted lines in (b) and (d) mark the chemical-accuracy scale.}
\label{Fig_UB_UNB_comparison}
\end{figure}

To quantify the contribution of the singly occupied $\ket{1}$ level, we compare the full $U_B+U_{\mathrm{NB}}$ ansatz with the $U_B$-only limit for LiH and H$_2$O. H$_2$ is not included because $U_B$ alone is sufficient in the relevant symmetry sector. In these $U_B$-only calculations, all non-bosonic broken-pair rotations $U_{\mathrm{NB}}$ are disabled and each active spatial orbital is restricted to the local states $\ket{0}$ and $\ket{2}$, yielding a compact pair-only representation analogous to a seniority-zero encoding with one qubit per spatial orbital. The molecular orbitals, Hamiltonian, and reference configuration are unchanged between the full $U_B+U_{\mathrm{NB}}$ and $U_B$-only calculations shown in Fig.~\ref{Fig_UB_UNB_comparison}.

For LiH, Fig.~\ref{Fig_UB_UNB_comparison} (a) shows the potential energy curves, while panel (b) gives the corresponding absolute errors relative to the FCI reference for both the $U_B$-only and $U_B+U_{\mathrm{NB}}$ spaces. In each case, the dashed lines denote exact diagonalization within the corresponding space, and the markers show the optimized VQE results obtained from the qutrit circuits. For both cases, the VQE results are essentially indistinguishable from their exact-space counterparts, with residual differences at or below the $10^{-5}~\mathrm{mHa}$ level. This agreement confirms that the singly occupied $\ket{1}$ level enables the additional correlations that distinguish the full qutrit space from the $U_B$-only space.

Near the potential energy minimum, the $U_B$-only calculation has an FCI error of approximately $5.34~\mathrm{mHa}$ at $D_{\mathrm{Li-H}}=1.5~\text{\AA}$, whereas the full $U_B+U_{\mathrm{NB}}$ result reduces this error to $1.35~\mathrm{mHa}$. Across the bond-length range, introducing the $\ket{1}$ level lowers the resulting energy by $2.23$--$74.82~\mathrm{mHa}$. The difference becomes particularly significant with bond stretching, where the pair-only model departs rapidly from FCI while the full qutrit circuit result continues to track the overall FCI potential-energy profile and remains below the chemical-accuracy scale over most of the range. The additional $\ket{1}$ configurations therefore recover essential broken-pair correlations that are inaccessible within the $U_B$-only pair space.

For H$_2$O, Fig.~\ref{Fig_UB_UNB_comparison} (c) and (d) show the potential energy curves and the corresponding errors relative to FCI for the $U_B$-only and $U_B+U_{\mathrm{NB}}$ encoded spaces. In the $U_B$-only case, the VQE result is indistinguishable from the exact diagonalization results for the pair-only space, with differences at the $10^{-3}~\mathrm{mHa}$ level. For the full $U_B+U_{\mathrm{NB}}$ case, exact diagonalization corresponds to the $\mathcal H_{90}$ reference discussed above, while the optimized VQE energy lies $0.125$--$2.83~\mathrm{mHa}$ above this reference across the sampled range. As shown in Fig.~\ref{Fig_UB_UNB_comparison}(c), this separation is small on the scale of the entire potential-energy curve and is of the same mHa order as the chemical-accuracy guideline.

Around the H$_2$O potential-energy minimum, at $D_{\mathrm{O-H}}=1.0~\text{\AA}$, the $U_B$-only space has an FCI deviation of $27.63~\mathrm{mHa}$. Introducing the singly occupied $\ket{1}$ level reduces the exact encoded-space error to $1.68~\mathrm{mHa}$, and the optimized full-qutrit VQE error is $1.92~\mathrm{mHa}$, both close to the chemical-accuracy scale. Over the sampled geometries, the additional $\ket{1}$ configurations lower the exact encoded-space energy by $9$--$170~\mathrm{mHa}$. Their contribution becomes especially large beyond the equilibrium region, where the energy difference between the pair-only and full qutrit spaces exceeds $100~\mathrm{mHa}$ and the $U_B$-only error grows rapidly with increasing O--H distance. The singly occupied level therefore provides an essential component of the H$_2$O correlation space absent from the pair-only encoding.

These comparisons establish a direct physical role for the third qutrit level. The singly occupied $\ket{1}$ state extends the compact pair-only encoding by admitting higher-seniority broken-pair configurations that become important upon bond stretching. The additional qutrit level therefore represents chemically relevant electronic configurations rather than serving only as a higher-dimensional carrier for an equivalent qubit register.

\section{Physical implementation and finite-error benchmarks}
\label{Sec:Implementation_and_error_benchmark}

Physical realization of the qutrit-native quantum simulation scheme requires a hardware platform with three well-resolved and coherently controllable energy levels for each physical unit, together with reliable initialization and measurement sequences. The platform should also have a high-fidelity qutrit gate set with general single-qutrit rotations and entangling two-qutrit operations to implement the logical operations used in this work. In this section, we discuss representative implementation strategies and finite-error benchmarks that offer order-of-magnitude fidelity guidelines for future experimental realizations. Since the present scheme is formulated at the logical level, the decomposition into elementary gates, circuit design, and connectivity strategy must be adapted to the native control capabilities and architecture of each hardware platform. 

Superconducting transmons provide a particularly direct platform for qutrit implementation because they behave as weakly anharmonic oscillators with readily addressable higher levels~\cite{goss2022high}. For example, randomized benchmarking of a five-qutrit superconducting processor yielded an average single-qutrit process fidelity above $99\%$ and a two-qutrit CSUM process fidelity of approximately $0.85$~\cite{morvan2021qutrit}. Additionally, a cross-Kerr interaction between two qutrits was used to implement entangling $CZ^\dagger$ and $CZ$ gates with process fidelities of $97.3(1)\%$ and $95.2(3)\%$, respectively~\cite{goss2022high}. Fast single-transmon qutrit Hadamard and $X$ gates with an average fidelity of $99.5\%$ have also been realized~\cite{yu2026efficient}. At the multi-qutrit level, the preparation of a three-qutrit GHZ-state fidelity of $0.951$ after error mitigation~\cite{goss2024extending} demonstrates combined use of local control, entangling gates, and qutrit-specific error mitigation. A universal qudit gate set and a decomposition scheme for synthesizing general qudit operations on transmons have also been proposed~\cite{fischer2023universal}, and multi-qutrit algorithms~\cite{blok2021quantum} and many-body simulations~\cite{ticea2025observation} have been reported. These results establish practical routes toward realizing the logical qutrit operation used in this work.

Molecular and semiconductor-defect spin qudits provide a complementary route in which multiple addressable levels arise naturally from high-spin electronic or nuclear manifolds. In an ensemble of $^{173}\mathrm{Yb}(\mathrm{trensal})$ molecules, transition-selective coherent control was used for a proof-of-concept quantum simulation~\cite{chicco2023proof} that exploits the multilevel nature of high-spin system. More recently, a qutrit quantum Fourier transform with full refocusing and complete state tomography was implemented on the same molecular platform, yielding fidelities of approximately $0.96$--$0.98$~\cite{rubin2026implementing}. These experiments establish transition-selective local qutrit control and readout on the molecular platform. Related solid-state experiments have encoded a logical qubit using a high-spin nuclear qudit in zinc oxide~\cite{lim2025demonstrating}, and electric-field control of a high-spin nuclear qudit system has also been reported~\cite{asaad2020coherent}. Both results indicate the capabilities of spin qudits in semiconductor hosts. While scalable coupling between distinct spin qudits remains less mature, a resonator-based architecture employing resonant photon exchange for two-qudit operations has been proposed and numerically analyzed~\cite{chiesa2023blueprint}.Electron-spin-mediated exchange coupling scheme~\cite{stemp2025scalable} provides another possible route toward scalable architectures based on electron or nuclear spin qudits.

Other high-dimensional platforms further broaden the implementation routes. Photonic qutrits have been extensively studied for their applications in quantum communication~\cite{vaziri2002experimental, luo2019quantum} and for foundational tests of quantum mechanics~\cite{lapkiewicz2011experimental}. Furthermore, a photonic variational eigensolver has experimentally encoded molecular computational spaces in the orbital-angular-momentum states of a single photon \cite{kim2024qudit}. Although this study used a $2^N$-dimensional qudit as a direct register for an $N$-qubit computational space, it demonstrated high-dimensional state preparation and projective measurement. Trapped-ion experiments have demonstrated a universal qudit processor using multiple internal ionic levels \cite{ringbauer2022universal}, while a separate theoretical study proposed trapped-ion-specific schemes for unitary-coupled-cluster variational calculations in a high-dimensional Hilbert space \cite{wang2025efficient}. Consequently, these studies provide additional experimental and theoretical routes for high-dimensional quantum simulation.

The studies discussed above show that the qutrit-native scheme can be implemented on several multilevel quantum platforms. Superconducting transmons are particularly advanced, with single- and two-qutrit gates, and multi-qutrit algorithms and simulations already reported. Although present qutrit gate fidelities remain below those of state-of-the-art qubit devices, these studies support experimental realization of the present scheme on near-term hardware. To provide quantitative guidance on the implementation fidelity required, we next test the two-qutrit H$_2$ simulation under finite errors.

\begin{figure}[!htbp]
\centering
\includegraphics[width=16cm]{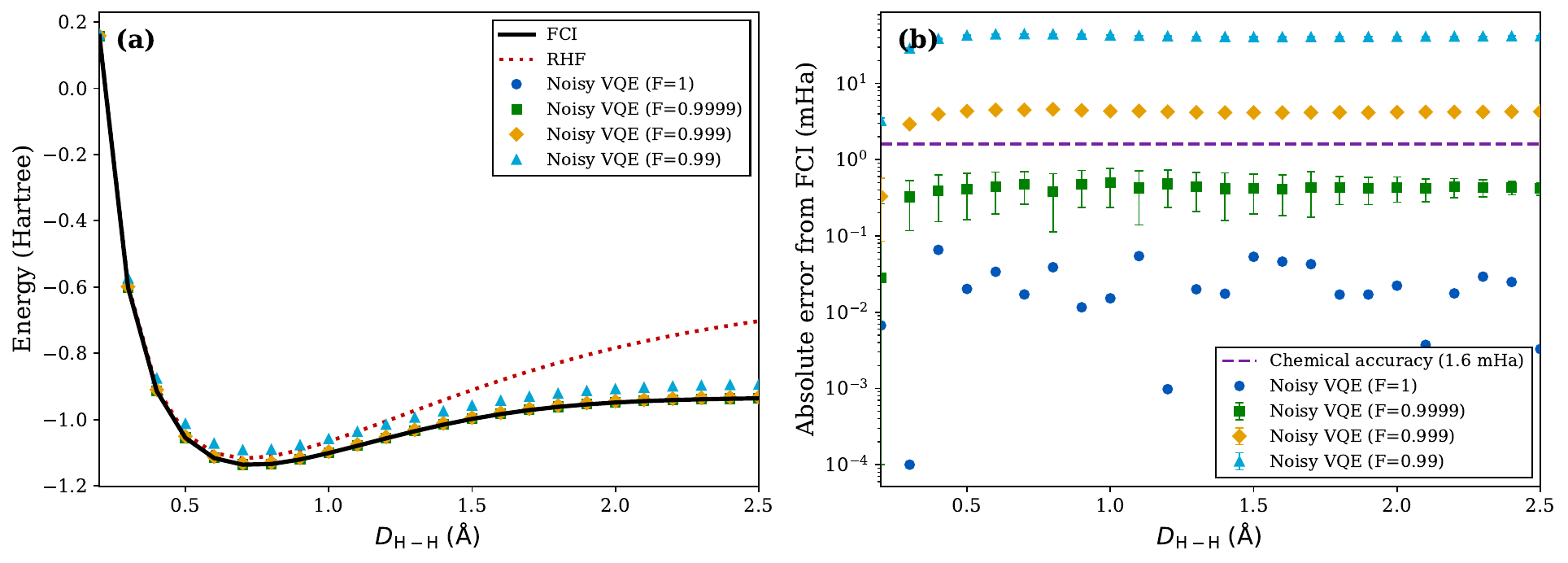}
\caption{Hardware-motivated finite-error benchmark for the two-qutrit H$_2$ VQE in the STO-3G basis. (a) RHF, FCI, and optimized qutrit-VQE potential-energy curves for effective fidelities $F=1$, $0.9999$, $0.999$, and $0.99$.
(b) Corresponding absolute VQE energy errors relative to FCI on a logarithmic scale. Error bars denote one standard deviation of the repeated final energy estimates. The horizontal dashed line marks the chemical-accuracy scale.}
\label{H2_sim_energy_curve_with_error}
\end{figure}

Because the specific error sources and mechanisms depend on the hardware platform, we model the overall implementation quality using a common effective fidelity $F$ applied to state preparation, the two-qutrit $U_B$ operation, the single-qutrit $U_P$ rotations used before measurement, and qutrit readout. This common fidelity is varied over $F=1$, $0.9999$, $0.999$, and $0.99$ to examine the resulting molecular energy accuracy. Each measurement setting uses $10^6$ shots, and each COBYLA objective value is obtained by averaging 100 independent noisy energy evaluations. The $F=1$ case is affected only by finite-shot sampling, unlike the ideal circuit results shown in Fig.~\ref{H2_sim_qutrit}.

Figure~\ref{H2_sim_energy_curve_with_error}(a) shows that the overall H$_2$ potential-energy profile remains well reproduced for effective infidelity scales of $10^{-4}$ and $10^{-3}$, while panel (b) shows the corresponding energy errors on a logarithmic scale. The sampling-only $F=1$ reference remains below $0.1~\mathrm{mHa}$ throughout the sampled range, and $F=0.9999$ gives sub-mHa errors that remain below the chemical-accuracy guideline at all geometries. At $F=0.999$, the errors increase to the few-mHa range, reaching approximately $4.6~\mathrm{mHa}$, but the shape of potential-energy curve and the sampled minimum remain well preserved. Although this fidelity level does not provide uniform chemical accuracy, the present two-qutrit benchmark shows that useful molecular energy simulations with few-mHa accuracy can be obtained at approximately the three-nine fidelity level. By contrast, $F=0.99$ produces errors reaching approximately $45~\mathrm{mHa}$, indicating that an overall implementation fidelity above $0.99$ is clearly required to avoid substantial quantitative degradation.

The various qutrit hardware developments discussed in this section support the feasibility of implementing the qutrit-native scheme on multilevel devices in the near term. The finite-error H$_2$ benchmark indicates that the three-nine fidelity level can support practical molecular simulations with reasonable accuracy. Because this benchmark uses only two qutrits and a single $U_B$ operation, larger molecular circuits will require systematic studies of how error accumulation and fidelity requirements depend on system size and circuit structure.

\section{Quantum-resource comparison}

\begin{figure}[!htbp]
\includegraphics[width=16cm]{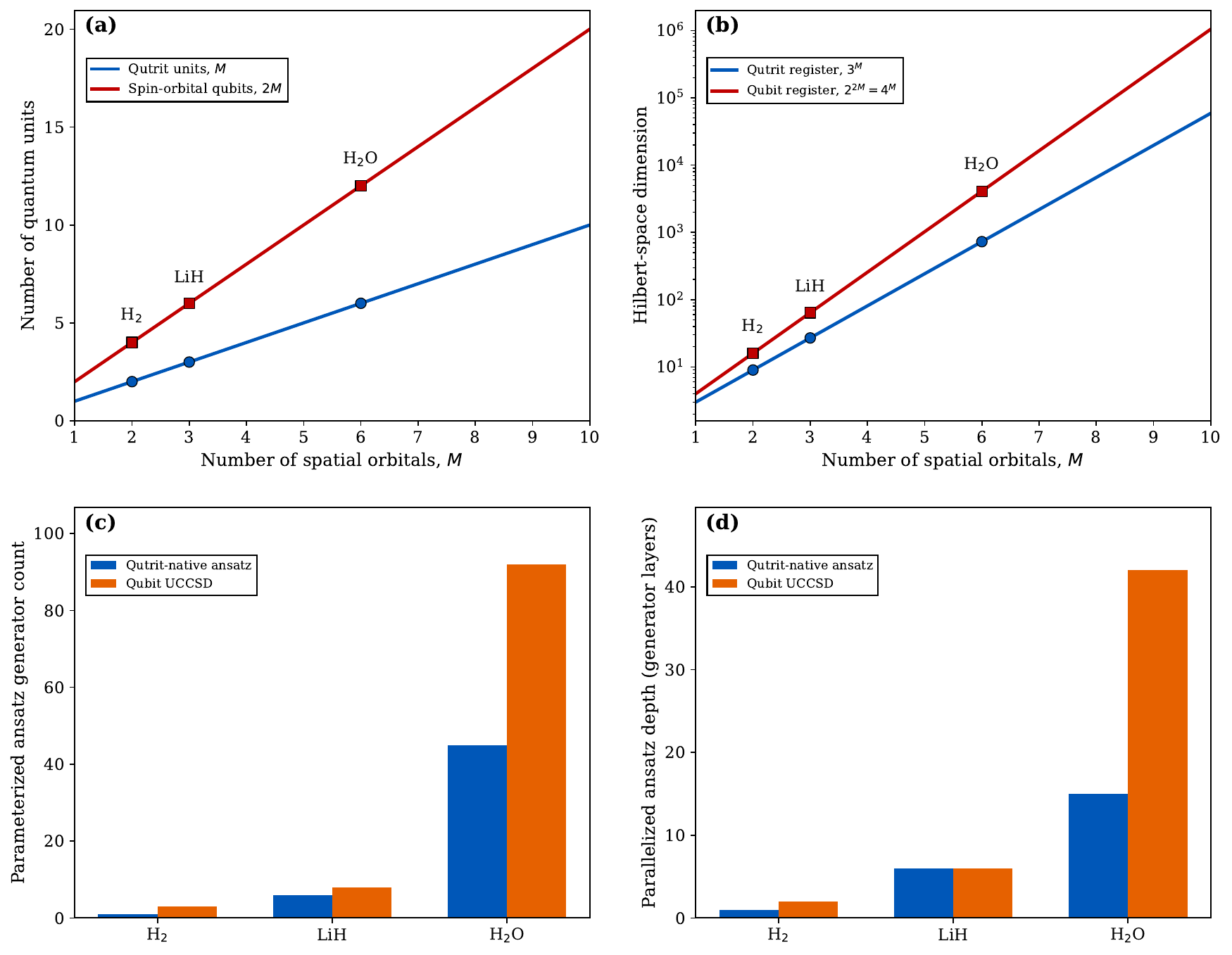}
\caption{Quantum-resource comparison between the qutrit-native scheme and the qubit-based UCCSD reference. (a) Numbers of quantum units, $M$ qutrits and $2M$ spin-orbital qubits. (b) Hilbert-space dimensions, $3^M$ and $4^M$. (c) Parameterized ansatz generator counts for H$_2$, LiH, and H$_2$O. (d) Corresponding parallelized ansatz depths in generator layers.}
\label{fig:resource_comparison}
\end{figure}

A central motivation for qutrit-native circuit design is to reduce the quantum resources required for quantum simulation. In this section, we quantitatively compare the resource requirements of the present qutrit scheme with those of a conventional qubit-based UCCSD ansatz. Fig.~\ref{fig:resource_comparison} compares the number of quantum units, Hilbert-space dimension, parameterized ansatz generator count, and parallelized ansatz depth for the active-space models considered in this work. The comparison shows an exponential reduction in Hilbert space dimension, and polynomial reductions in ansatz generator count and parallelized ansatz depth.

At the quantum unit level, one qutrit is assigned to each active spatial orbital in the present scheme. An active space containing $M$ spatial orbitals therefore requires $M$ qutrits, whereas a conventional spin-orbital representation uses $2M$ qubits. As shown in Fig.~\ref{fig:resource_comparison}(a), the H$_2$, LiH, and H$_2$O models considered here consequently use $2$, $3$, and $6$ qutrits instead of $4$, $6$, and $12$ spin-orbital qubits, respectively. The corresponding Hilbert-space dimensions are $3^M$ and $2^{2M}=4^M$ as shown in Fig.~\ref{fig:resource_comparison}(b), and their ratio grows as $(4/3)^M$.  

These dimensions refer to the full Hilbert spaces before particle-number and spin constraints are imposed. The fixed-electron-number qutrit occupation spaces for H$_2$, LiH, and H$_2$O contain $3$, $6$, and $90$ states, respectively. For H$_2$ and LiH, these spaces coincide with the complete singlet spaces. For H$_2$O, the complete singlet CASCI space is $105$-dimensional, whereas the qutrit encoding spans a $90$-dimensional subspace selected by the pairing rule discussed above. The effect of the $15$ omitted singlet states arising from the second independent singlet coupling is examined explicitly in Appendix~\ref{SI_chapter_B}.

We next compare the variational circuit structures in terms of the parameterized ansatz generator count. As shown in Fig.~\ref{fig:resource_comparison}(c), the qutrit circuits used for H$_2$, LiH, and H$_2$O contain $1$, $6$, and $45$ generators, respectively. For the qubit reference, we use a conventional spin-orbital UCCSD ansatz in the same active spaces. For the closed-shell references considered here, if $o$ and $v$ denote the numbers of occupied and virtual spatial orbitals, respectively, the corresponding generator counts are
\begin{equation}
N_{\mathrm S}=2ov,
\qquad
N_{\mathrm D}=2\binom{o}{2}\binom{v}{2}+o^2v^2.
\label{eq:uccsd_generator_count}
\end{equation}
This gives total UCCSD generator counts of $3$, $8$, and $92$ for H$_2$, LiH, and H$_2$O, respectively. These counts agree with the UCCSD parameter counts reported for the corresponding molecular benchmarks in
Ref.~\cite{li2021software}. Therefore, the qutrit ansatz uses fewer parameterized generators and correspondingly fewer parameterized gate blocks than the qubit-based UCCSD scheme for all three molecules.

Panel (d) further compares the parallelized ansatz depth. Each parameterized generator is treated as one basic logical operation, all-to-all connectivity is assumed, and generators acting on disjoint quantum units are assigned to the same layer. Under this scheduling, the parallelized circuit depths of the qutrit and qubit ansatzes are $1$ and $2$ for H$_2$, $6$ and $6$ for LiH, and $15$ and $42$ for H$_2$O, respectively. 

For a single all-pair sweep of the present qutrit construction, assigning one $U_B$ generator and up to two directed $U_{\mathrm{NB}}$ generators to each unordered orbital pair gives an upper bound on the generator count,
\begin{equation}
N_{\mathrm{gen}}^{(\mathrm{qutrit})}
\leq
3\binom{M}{2},
\label{eq:qutrit_generator_upper_bound}
\end{equation}
which scales quadratically, $\mathcal{O}(M^2)$, with $M$. By comparison, the conventional spin-orbital UCCSD generator count in Eq.~\eqref{eq:uccsd_generator_count} scales quartically as $\mathcal{O}(M^4)$ at fixed filling. 

This section quantitatively compared the resource requirements of the qutrit and qubit approaches in terms of quantum units, Hilbert-space dimension, and variational circuit structure. Across these measures, the qutrit-native construction uses fewer quantum units, a smaller Hilbert space, fewer parameterized ansatz generators, and equal or lower parallelized ansatz depths than the qubit-based UCCSD reference. Although the decomposition of the logical ansatz blocks into elementary gates can vary with the hardware and implementation strategy~\cite{romero2019strategies, yordanov2020efficient, kuhn2019accuracy}, the qutrit construction consistently reduces the ansatz generator count and parallelized depth in the present comparison.

\section{Conclusion and future work}

We have developed a qutrit-native framework for molecular electronic-structure simulation that maps each active spatial orbital to one qutrit and uses the $\ket{1}$ level to encode singly occupied, broken-pair configurations. The framework provides the corresponding ansatz generators and a projected-Hamiltonian energy estimator based on qutrit populations and coherences. Benchmarks for H$_2$, LiH, and H$_2$O validate the framework across
correlation spaces of increasing size and complexity. Comparisons with the pair-only restriction show that the singly occupied $\ket{1}$ level recovers broken-pair correlations that become increasingly important upon bond stretching. In terms of quantum resources, the encoding uses $M$ qutrits rather than $2M$ spin-orbital qubits. The parameterized ansatz generator count of the present qutrit construction scales as $\mathcal O(M^2)$, compared with $\mathcal O(M^4)$ for conventional spin-orbital UCCSD at fixed filling. For H$_2$O, the compact 90-state encoding retains agreement with FCI at the few-mHa level despite the partial spin-coupling truncation and preserves the overall potential-energy profile across the bond-stretching range. Together, these results show that reductions in register and ansatz resources can coexist with few-mHa electronic-structure accuracy in the present multi-electron benchmark.

Physical implementation of the present scheme requires coherent single-qutrit control, entangling two-qutrit operations, and reliable initialization and readout. The two-qutrit H$_2$ finite-error benchmark indicates an effective fidelity
scale around the three-nine level for retaining useful molecular-energy accuracy. These results provide an initial quantitative fidelity guideline for experimental implementations of the qutrit-native scheme. As qutrit devices and multilevel gate techniques are currently under active investigation, these results suggest that experimental realization may become possible on near-term multilevel quantum devices. More broadly, these results motivate qudit-native algorithm design in which additional local levels represent physically meaningful sectors of the simulated problem rather than serving only as additional computational capacity. Extensions to larger active spaces, more flexible ansatz and spin-coupling schemes, and general multilevel architectures are natural directions for further study.

\begin{acknowledgments}
This project was supported by the National Research Foundation of Korea under Grant No. RS-2023-00256050.
\end{acknowledgments}


\bibliography{qutrit_quantum_chem_sim}



\appendix

\setcounter{figure}{0}
\renewcommand{\figurename}{Fig.}
\renewcommand{\thefigure}{A\arabic{figure}}
\renewcommand{\theHfigure}{A\arabic{figure}}

\section{Molecular orbital selection of LiH}
\label{SI_chapter_A}

In Appendix A, we describe the molecular-orbital selection used for the LiH simulation and quantify the associated active-space error. To separate the active-space truncation error from other qutrit-simulation errors, we first compare several classical electronic-structure models of LiH. In the STO-3G basis, LiH has six spatial molecular orbitals.  The FCI calculation correlates all four electrons in this orbital space, whereas the CASCI calculations keep the lowest Li $1s$-like molecular orbital doubly occupied as a frozen core. CASCI(2,5) therefore retains the two valence electrons in all five remaining spatial orbitals and provides the frozen-core active-space reference.  

\begin{figure}[!htbp]
\includegraphics[width=16cm]{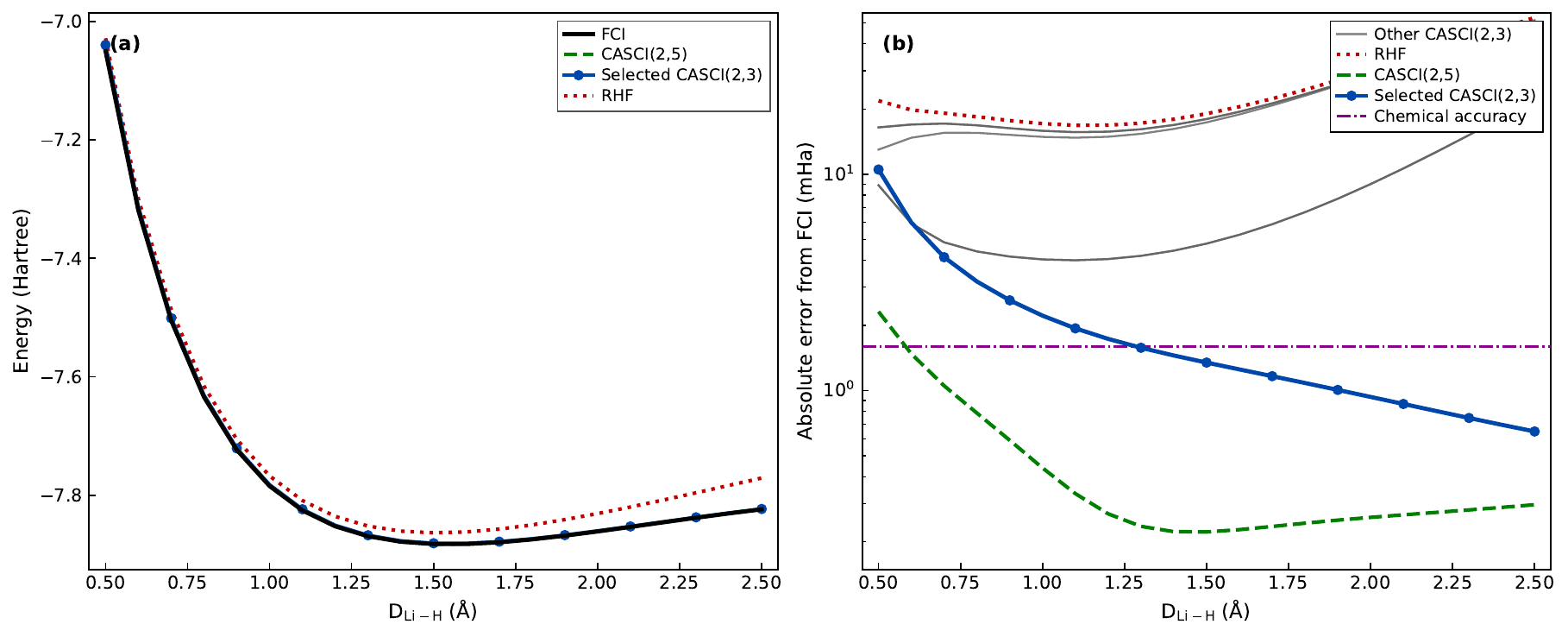}
\caption{Active-space truncation for LiH. (a) Potential-energy curves from RHF, FCI, frozen-core CASCI(2,5), and the selected CASCI(2,3) active space. (b) Corresponding absolute errors relative to FCI on a logarithmic scale. The
gray curves show alternative fixed CASCI(2,3) orbital subsets within the plotted range. The horizontal dash-dotted line marks the chemical-accuracy scale.}
\label{FigS1_lih_active_space_selection}
\end{figure}

Within the occupation encoding employed in this work, however, this active space would require five qutrits. Because a central goal of the present scheme is to reduce quantum-resource requirements while retaining electronic-structure accuracy, we further reduce the active space to three spatial orbitals, corresponding to three qutrits. We therefore considered all fixed three-orbital subsets of the five-orbital valence space. The active space selected for the main-text simulation consists of orbitals with dominant H $1s$, Li $2s$, and Li $2p_x$ character, where the $x$ direction is chosen along the Li--H bond. As shown in Fig.~\ref{FigS1_lih_active_space_selection}(a), the resulting CASCI(2,3) potential-energy curve closely follows the FCI curve throughout the chemically relevant region despite the reduction from five to three active spatial orbitals.

Figure~\ref{FigS1_lih_active_space_selection}(b) quantifies the corresponding active-space errors. The frozen-core CASCI(2,5) result lies well within chemical accuracy around the equilibrium geometry. Although most fixed
CASCI(2,3) choices exhibit substantially larger deviations, the selected three-orbital space used in the main text (blue line) remains within the $1~\mathrm{kcal\,mol^{-1}}$ threshold near equilibrium, with errors of approximately $1.5~\mathrm{mHa}$ at both $D_{\mathrm{Li-H}}=1.5$ and $1.6~\text{\AA}$. Its deviation increases primarily in the strongly compressed region. Among the nine alternative three-orbital subsets, five lie within the plotted range and are shown by the gray curves, with symmetry-related curves overlapping, while the remaining four have errors of order $10^2~\mathrm{mHa}$ and fall outside the vertical range of panel (b). This behavior makes the selected space a favorable compromise between active-space accuracy and qutrit resource requirements.

\section{Spin-coupling truncation for four singly occupied orbitals in H$_2$O }
\label{SI_chapter_B}

\begin{figure}[!htbp]
\includegraphics[width=8cm]{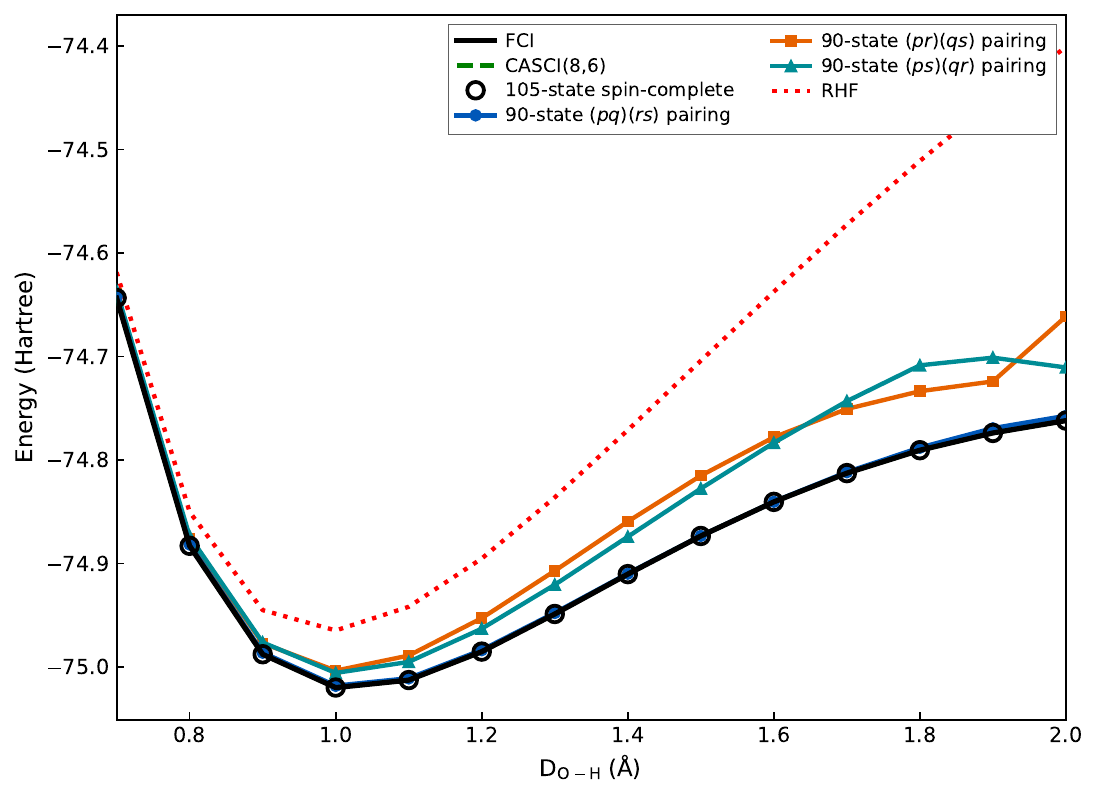}
\caption{Potential-energy curves of H$_2$O in the STO-3G basis at a fixed H--O--H angle of $104.5^\circ$. Results are shown for RHF, FCI, frozen-core CASCI(8,6), the three 90-state spaces defined by the $(pq)(rs)$,
$(pr)(qs)$, and $(ps)(qr)$ pairing rules, and the 105-state spin-complete space. The hollow circles denote the 105-state result, which overlaps the CASCI(8,6) curve.}
\label{H2O_appendix_singlet_pairing_energy_curves}
\end{figure}

\begin{figure}[!htbp]
\includegraphics[width=16cm]{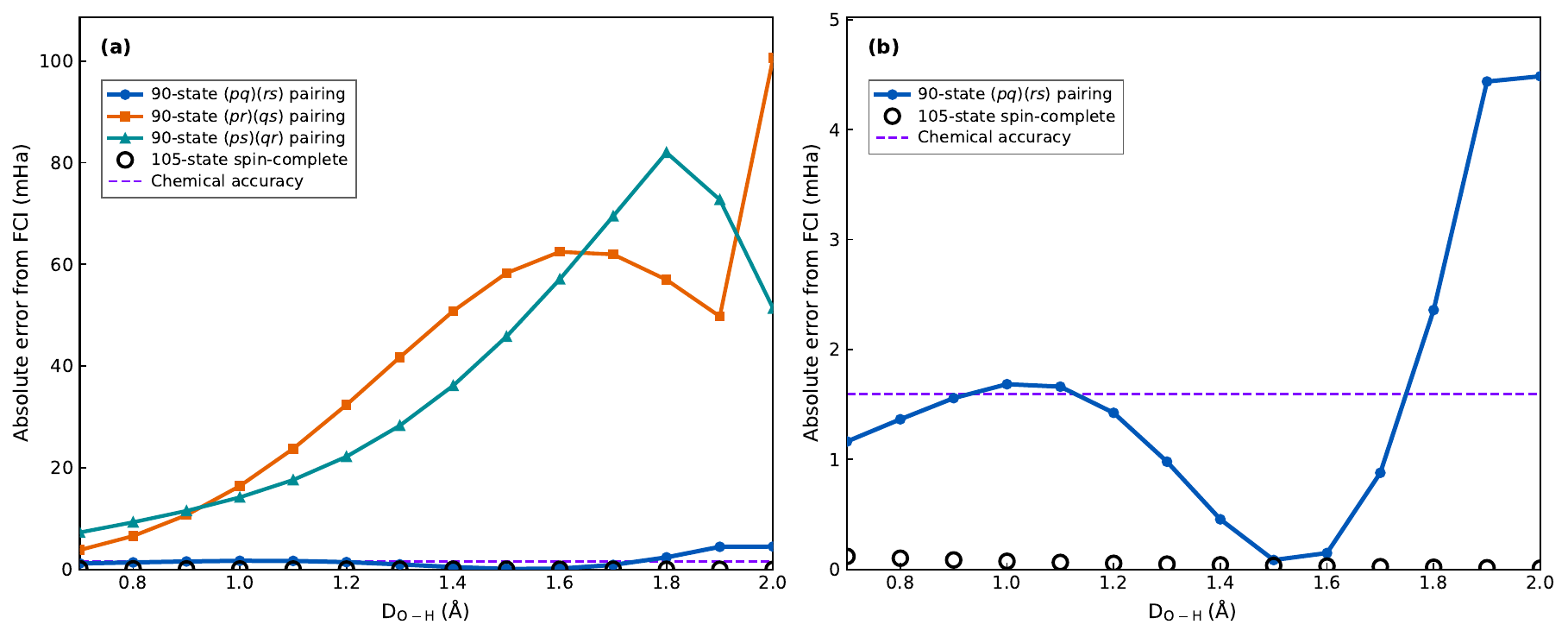}
\caption{Absolute energy deviations from FCI for the H$_2$O pairing-rule comparison shown in Fig.~\ref{H2O_appendix_singlet_pairing_energy_curves}. (a) Errors for the three 90-state spaces defined by the $(pq)(rs)$, $(pr)(qs)$, and $(ps)(qr)$ pairing rules, together with the 105-state spin-complete result. (b) Enlarged view of the selected $(pq)(rs)$ space and the 105-state result. The horizontal dashed line marks the chemical-accuracy scale.}
\label{H2O_appendix_singlet_pairing_errors}
\end{figure}

In Appendix B, we quantitatively analyze the truncation introduced by the qutrit occupation encoding for configurations containing four singly occupied spatial orbitals. The comparison is performed for H$_2$O in the STO-3G basis at a fixed H--O--H angle of $104.5^\circ$, using the same frozen-core CASCI(8,6) model as in the main text. Freezing the O $1s$-like molecular orbital leaves eight active electrons distributed among six spatial
orbitals. As shown in Fig.~\ref{H2O_appendix_singlet_pairing_energy_curves}, CASCI(8,6) closely reproduces the FCI potential-energy curve. Its deviation from FCI is only approximately $0.011$--$0.118~\mathrm{mHa}$ over the bond-length range considered here. CASCI(8,6) therefore provides an accurate active-space representation. To maintain a common benchmark throughout this work, all errors in Fig.~\ref{H2O_appendix_singlet_pairing_errors} are reported relative to FCI.

For eight electrons in six qutrits, the allowed occupation strings satisfy
\begin{equation}
    n_i\in\{0,1,2\},
    \qquad
    \sum_{i=1}^{6}n_i=8,
\end{equation}
and form a 90-dimensional occupation space. The complete spin-adapted singlet space of the same CASCI(8,6) problem instead contains 105 configuration-state functions (CSFs). The difference arises entirely from
occupation patterns containing four singly occupied orbitals. There are $\binom{6}{4}=15$ such patterns. For each pattern, four spin-$1/2$ electrons support two linearly independent total-singlet spin functions,
whereas the qutrit occupation string retains only one prescribed singlet coupling. Consequently, restoring the second singlet CSF for each of the 15 four-open-shell patterns increases the dimension from 90 to 105, recovering
the complete singlet CASCI(8,6) space. The corresponding 105-state result coincides with the CASCI(8,6) energy in Fig.~\ref{H2O_appendix_singlet_pairing_energy_curves}.

To specify the retained spin function, let $p<q<r<s$ denote the four singly occupied orbitals ordered by their molecular-orbital energies.  The three pairwise singlet-coupling patterns may be denoted by
\begin{equation}
    (pq)(rs),\qquad (pr)(qs),\qquad (ps)(qr),
\end{equation}
where
\begin{equation}
    \ket{S_{pq}}
    =
    \frac{1}{\sqrt{2}}
    \left(
        \ket{\alpha_p\beta_q}
        -
        \ket{\beta_p\alpha_q}
    \right)
\end{equation}
defines a singlet-coupled pair.  These three valence-bond-like pairings are not linearly independent and together span the two-dimensional singlet space of the four open-shell electrons.  Applying any one of these pairing rules to the four-open-shell occupation patterns therefore defines a corresponding 90-dimensional truncation of the 105-dimensional singlet-complete space~\cite{szabo2012modern}.

In the simulations presented in the main text, we retain the $(pq)(rs)$ coupling, which singlet-couples the two lower-energy singly occupied orbitals and, separately, the two higher-energy orbitals. This choice
provides a simple and deterministic orbital-ordering rule while preserving a compact pairwise spin-correlation structure. Its broader physical motivation is supported by spin-coupled generalized-valence-bond analyses,
which have shown that molecular ground states near equilibrium are often dominated by a compact perfect-pairing spin function when the active orbitals are organized into chemically interacting pairs~\cite{dunning2016insights}.  The identity and relative importance of the dominant spin coupling, however, can depend on the molecular system, orbital character, and molecular geometry~\cite{xu2015generalized}. 

Because the present qutrit encoding associates each four-open-shell occupation string with only one retained spin function, an unambiguous and resource-efficient mapping requires a fixed selection rule. We therefore adopt the $(pq)(rs)$ pairing as a deterministic default choice within the present framework. The H$_2$O comparison below provides a direct quantitative test of this choice. Among the three possible pairing rules, the $(pq)(rs)$ choice consistently gives the smallest FCI deviation and maintains errors within a few $\mathrm{mHa}$ throughout the bond-length range considered.

Figure~\ref{H2O_appendix_singlet_pairing_errors} compares the three possible pairing rules. For H$_2$O, the selected $(pq)(rs)$ coupling gives the lowest energy and hence the smallest deviation from FCI, at every
geometry examined. Its FCI error ranges from approximately $0.085$ to $4.49~\mathrm{mHa}$ over $D_{\mathrm{O-H}}=0.7$--$2.0~\text{\AA}$. Around the equilibrium region, the error is approximately $1.56$--$1.68~\mathrm{mHa}$ for $D_{\mathrm{O-H}}=0.9$--$1.1~\text{\AA}$, comparable to the chemical-accuracy scale. By contrast, the $(pr)(qs)$ and $(ps)(qr)$ rules produce errors of several to tens of $\mathrm{mHa}$ and become substantially less accurate as the O--H bonds are stretched.

These results show that the selected 90-state space provides a favorable compromise between electronic-structure accuracy and quantum-resource cost for the H$_2$O model considered here. The 15-dimensional reduction from the
spin-complete CASCI(8,6) space is intrinsic to the present six-qutrit occupation encoding, while the $(pq)(rs)$ rule keeps the associated error within a few $\mathrm{mHa}$ throughout the tested geometries.

\end{document}